# Teaching the Action Principle in Optics

Refath Bari

July 19, 2021

Abstract

The Principle of Least Action (PLA) in Optics can be confusing to students, in part due to the Calculus of Variations, but also because of the subtleties of the actual principle. To address this problem, three simulations of the PLA are presented so students can learn the Action Principle in an experiential and interactive manner. Simulations such as MIT's OpenRelativity and PhET's Quantum Mechanics have become a popular pedagogical tool to demystify abstract physical phenomena. This paper aims to help undergraduate students understand the Action Principle by introducing three numerical simulations: light reflecting in equal angles, light refracting in different mediums, and light moving between two points in the least time. The interactive simulations discussed in this paper are available [here](#).

**1 Introduction**
The Action Principle can be difficult for students, not only due to its abstract nature but also because it is typically introduced analytically, not visually. The action itself can be hard to digest, given its intangible nature, unlike velocity or even acceleration which can be understood intuitively. Students may be confused why the Lagrangian must be T-V instead of T+V, as mechanical energy is [5]. They may be confused by the abstraction of "nearby paths" and wonder why light can't simply travel in straight lines to minimize distance [20]. Furthermore, Fermat's Principle and the Calculus of Variations (COV) is not about minimizing a variable, but a function, which can be difficult for students to comprehend [10][17] —especially when applied to the nearby paths of the Principle of Least Action (PLA). Students are not given appropriate visuals to understand COV or PLA, thus exacerbating their confusion and misconceptions.

In light of these difficulties, numerical simulations may help students understand the PLA. Interactive Simulations have proven to be useful tools in teaching physics to undergraduate students, particularly in distilling abstract ideas into tangible ones. For instance, OpenRelativity [27], a product of MIT's GameLab, has used mouse and keyboard movements to demonstrate time dilation and length contraction visually. PhET's Quantum Mechanics (QM) simulations [23] have tied students' actions in the program to animations that explain QM. However, few simulations have accounted for the intersection of these two fields -- the PLA [19]. The PLA motivates the relativistic principle of following the world-line of maximal aging as well as the quantum mechanical principle of exploring all paths. This paper takes a slightly different approach in that it encourages students to engage in the actual creation of the programs as opposed to simply interacting with pre-written simulations. Although students will benefit either way, the latter method will motivate students to understand the mechanics of the PLA as they implement it themselves. [21] [22]

The purpose of this paper is to present three numerical simulations of the Action Principle in Optics. In doing so, it will leave students with an intuitive understanding of the PLA. These simulations would be suitable for an Optics or Advanced Mechanics course that deals with the Action principle, reflection, refraction, or the Brachistochrone problem. The first simulation is light reflecting in equal angles off a metal. The second simulation involves light refracting through different mediums. The final simulation involves the Brachistochrone problem, solved using Johann Bernoulli's application of Snell's Law to the problem, treating a particle as a beam of light [9].

## 2 Simulations

The author chose to write the programs in Python to make it accessible to the broadest student audience possible. Python is an easy-to-learn language and even students who are not familiar with programming can learn it easily [14]. Python is consistently ranked one of the most popular and beginner-friendly programming languages available [12]. In making the Action Principle as intuitive as possible, the author chose the 3D graphics module VPython, which is built on the Python language so that students can interact with simulations in a navigable 3d-space. As it is based on Python, VPython is easy for students to master (i.e., the simple command `box()` creates a 3D box with default lighting and camera-positioning parameters. It also activates mouse interactions so the user can rotate, zoom, and move through the 3D space) VPython has been growing in popularity as a pedagogical tool to teach physics in an intuitive, interactive manner [4][26]. Even though Reflection, Refraction, and the Brachistochrone are mathematically trivial, their formulas or proofs in isolation don't give students insight into the nature of the physical phenomena, unlike a simulation [13] [24]. As opposed to a textbook, students can actually interact with a simulation, experiment with parameters, and analyze real-time graphs. This affords students an experiential learning experience with a textbook [3][18]. As opposed to passive learning from a textbook, simulations encourage students to be active learners by changing parameters, modifying variables, and interacting with the visualizations [1][15]. All code for the simulations have been made available through easily-customizable Jupyter notebooks, in which teachers can view and modify the code, students can interact with the simulations, and instructors can view tutorials for using the programs. To create these programs themselves, students simply click on the link above, open a notebook, and start coding.

Many of the geometric proofs for the three phenomena below are elementary, requiring no more than high school geometry. In addition, the examples of reflection, refraction, and the

Brachistochrone problem to demonstrate the PLA have been extensively covered in undergraduate textbooks. So why the need for simulations? Simply put, simulations have the potential to catalyze discovery and engagement in the physics classroom in a way that no proof or textbook can [7][11][25]. In fact, it has been demonstrated that teaching reflection, refraction, and optics via an experiential, visual, and interactive approach using programming benefits students by fostering active learning and trains students to approach problem-solving through a numerical approach for traditionally analytic problems. [2] [16] [26]

*2.1 Reflection*

The Law of Reflection states that light will reflect off surfaces such that $\theta_i = \theta_r$, where $\theta_i$ and $\theta_r$ are the angles of incidence and reflection, respectively. The first simulation demonstrates the law of reflection by plotting all possible trajectories that light can take between two points if it must strike a metal surface. The first simulation serves three pedagogical purposes in helping students understand the action principle as applied to the law of reflection: help students verify $\theta_i = \theta_r$, give insight into the optimization proof for reflection, and resolve any student misconceptions.

First, the simulation can aid students in verifying or deriving $\theta_i = \theta_r$. Once informed that light takes the path of shortest time or least action, students can then derive $\theta_i = \theta_r$ in two ways. Observing the Light Time [s] against the Angle [deg.] plot, students can recognize that the light time is minimized when the blue (incident angle) and green (reflected angle) plots intersect. Alternatively, students can observe the Action plotted against the ratio Incident:Reflected, and note that the action is minimized when the ratio is 1 (i.e., the incident and reflected angles are equal).

Fig. 1: Students will recognize that the light time approaches a minimum as the incident and reflected angles converge.

Second, the simulation can give students insight into the optimization proof that results in Reflection. By programming the simulation themselves instead of using a pre-written PhET simulation, students will be able to implement a differential equation whose solution is the reflected light path. Students will implement $\frac{\sqrt{a^2+x^2}+\sqrt{b^2+(L-x)^2}}{v_1}$, where *a* and *b* are the heights of the endpoints, *x* is the hit-point where light hits the metal, and *L* is the horizontal distance between the endpoints. By programming the optimization function and incrementally changing the parameters (i.e., locations of the endpoints), students will be able to internalize why the action principle dictates that light must reflect such that $\theta_i = \theta_r$.

Third, the simulation can clarify student misconceptions. Despite the trivial nature of the reflection formula, students are still liable to misconceptions or questions, especially because the mathematical method of measuring symmetric distances from the normal is neither informative nor gives insight into the phenomenon of reflection [24]. Furthermore, students may inquire why it isn't faster to travel straight to the metal bar and then reach the end point, or other questions regarding the nature of the least-time path. All these inquiries may be addressed via the simulation.

To emphasize that light takes the path of least action, all possible paths between the endpoints are drawn, and the least-action path is bolded in red. The challenge in programming this feature is for students to recognize that a `for` loop should be used to iterate the horizontal position of the light ray. Afterwards, the `diff_angle` method can be used to find $\theta_i$ and $\theta_r$.

```
for i in range(int(v1.x),int(v3.x)):
        v2.x = i, c = curve([v1,v2]), c.append(pos = v3)
        incident = degrees(diff_angle(incRef,v2-v1))
        reflected = degrees(pi - diff_angle(incRef,v3-v2))
```

The user is able to click to denote the initial and final position of the beam of light. This interactive element reinforces the different initial values given for Classical Newtonian Mechanics rather than the Action Principle. In the former, we are given the initial position and velocity. In the latter, we are simply given boundary conditions — the endpoints of the light beam's trajectory. As all possible trajectories from A to B are traced out on the screen, real-time plots are generated indicating the relationship between the Incidence Angle, Reflection Angle, Light Time, and Action. This element enables the student to conjecture by inspection that the Action is at a minimum when the Light Time is minimized and the Reflection and Incidence Angles are equal.

*2.2 Refraction*

The Law of Refraction states that the angle of incidence and refraction for a light ray changing mediums is governed by $n_1 sin\theta_1 = n_2 sin\theta_2$, where $n_1$ and $n_2$ are the refractive indices of the two mediums and $\theta_1$ and $\theta_2$ are the angles of incidence and refraction, respectively. The second simulation demonstrates the Law of Refraction by showing all possible trajectories for light to refract between two endpoints selected by the user, where the endpoints are in different mediums. It serves three pedagogical purposes in helping students understand refraction: help students verify the law of refraction, give insight into the optimization function, and clarify student questions about how the PLA dictates the law of refraction.

First, the simulation enables students to verify or derive $n_1 \sin\theta_1 = n_2 \sin\theta_2$. The experimental derivation of Snell's Law defied many of the greatest philosophers and physicists of the past, including Ptolemy and Kepler. So how can we expect students to derive it in an hour? In fact, this was precisely the question that Vaughn et al. posed in 1975, and to their happy surprise, found that removing lab reports and instead challenging students to derive Snell's law by examining the experimental data transformed the spirit of the classroom from menial lab work to exciting scientific research [20]. The refraction simulation can perform a similar function in the classroom by catalyzing student engagement in the Principle of Least Action.

Fig. 2: After multiple trials, students will note that the incident angle is always greater than the refracted angle due to light's slower speed in the denser medium. Students can also increase the argument of the refraction function and note that the refracted angle approaches closer to the normal.

This simulation comes with two graphs and several parameters available for students to change. Students can observe when the light time is minimized in the plot of Light Time [s] against the Angle [deg.] They can also note that the Action is minimized on the path in which the ratio $\sin\theta_1 : \sin\theta_2$ is equal to the ratio $n_1 : n_2$. Students can also try experimenting by changing parameters such as the refractive index of the medium that light enters, as well as the initial start and end positions of the light ray. By tuning and experimenting with all these parameters and observing the graphs, students will be able to -- at the very least -- verify Snell's Law. However, given enough time, much like Vaughn et al.'s pedagogical experiment, the refraction simulation may be able to engender an air of scientific curiosity in the classroom and help students experientially discover what the PLA dictates the law of refraction to be.

Similar to the first simulation, the second simulation can also give students insight into the optimization proof that results in Refraction. Once again, by programming the simulation themselves, students will no longer need to afford benefit of the doubt to the optimization function $\frac{\sqrt{x^2+z^2}}{v_1} + \frac{\sqrt{(d-z)^2+y^2}}{v_2}$ that results in refraction, where $d$ is the horizontal distance between the endpoints, $z$ is the location of the hitpoint where light hits the second medium, $x$ and $y$ are the vertical heights of the two endpoints, and $v_1$ & $v_2$ are the velocities of light in the two mediums. Instead, students will have the opportunity to program the function into the code themselves, and understand all the parameters involved and internalize the refractive behavior of light.

Third, the Refraction simulation can clarify student questions. Students may inquire why the action is not minimized by (A) a straight path through the water or (B) a path that minimizes the distance traveled in the slower medium. In this case, students can deduce the answer independently by observing the graphs for Path A and B, which show the light time and Action are not minimized. Students may also inquire how the refractive index of the medium *n* affects $\theta_i$ and $\theta_r$. Upon changing that parameter, students can observe from the graph that as *n* increases, $\theta_i$ and $\theta_r$ diverge and $\theta_r$ approaches closer to the normal. There are no particularly challenging components of the refraction code, as its architecture is similar to the reflection simulation.

Like the previous program, the user can click to denote the initial and final positions of the beam of light. As every possible trajectory is traced out, this reinforces light's tendency to follow the PLA, examine every path and select the one of least action, as opposed to the frame-by-frame nature of Classical Newtonian Mechanics, where position and momentum are given for the present and asked to be found for some time in the future. Once again, real-time plots of the Incidence Angle, Refraction Angle, Light Time, and Action are traced. This element not only enables the student to identify the path of least action, but also thereby conjecture Snell's Law.

*2.3 Brachistochrone*

Consider a beam of light traveling through an inhomogeneous medium whose index of refraction $n \propto \frac{1}{\sqrt{h}}$, where *h* is the height between the medium interface and light ray. The light ray will pursue the path of least action, but what is that path? This optics problem is equivalent to the Brachistochrone problem, posed by Johann Bernoulli in 1696: "Given two points A and B in a vertical plane, what is the curve traced out by a point acted on only by gravity, which starts at A and reaches B in the shortest time?" This is the defining problem for Calculus of Variations, the theory behind the Principle of Least Action. As such, exposing students to the Brachistochrone problem will enable them to better understand the optical properties of light as dictated by the PLA. Indeed, Bernoulli's solution to the problem was to exploit that phenomenon which naturally chose the path of least time, thus reframing a mechanics problem as an optimization problem for light. The third simulation demonstrates this problem by enabling the student to draw various paths between two endpoints and conjecture which paths have the least action, based on graphs of the velocity, time, Action, Kinetic Energy, and Potential Energy. The third simulation serves three pedagogical purposes in helping students understand the PLA as applied to the Brachistochrone problem: an appreciation for the competition between path length and acceleration, insight into how the PLA applies to the Brachistochrone problem, and the analogy between light and a ball rolling down a cycloid.

First, the simulation can give students an insight into the tug-of-war between the two parameters of path length and acceleration that determine the path of least time. In programming the Brachistochrone simulation, students will note that the velocity $v$ is simply a function of height $\sqrt{2gh}$, a consequence of energy conservation [9].

Second, the simulation can demonstrate how the principle of least action determines the path of least time in the Brachistochrone problem. Students can observe the action for different paths in the Brachistochrone simulation. By experimenting with different paths and tinkering with the boundary conditions, students can understand why a straight line or a circular arc are not the least-time solutions [6].

Last, the third simulation can also aid students in recognizing the analogy between the Brachistochrone problem and the behavior of light. The goal of the problem is to find the path of least time between two endpoints in a gravitational field. Johann Bernoulli cleverly exploited a phenomenon that naturally takes the path of least time: light. In fact, Bernoulli's solution to the problem was to imagine the Brachistochrone as a path of light traveling through an optically homogeneous medium. Much like the Brachistochrone problem, the behavior of light as it travels through different mediums is also a competition between two parameters, namely a balance between minimal time spent in the slower medium and a shorter path. Bernoulli's solution to the Brachistochrone problem involved imagining light moving through many mediums of different refractive indices. Light would naturally take the path of least time, thus solving the problem. Since light follows Snell's Law when changing mediums, whatever path light takes would have to satisfy Snell's Law at each point.

Fig. 3: Students may experiment with a variety of paths to find the path of least time.

The Brachistochrone simulation is naturally challenging to engineer, due to the many degrees of freedom (i.e., angle and number of curve segments) afforded to the user. The key idea is to create a condition which helps the program recognize if the particle has finished traveling each segment of the curve. The second part of the `while` loop condition is to check whether the second endpoint is below the first one, so as not to violate energy conservation. Students then implement the velocity as a function of height $\sqrt{2gh}$.

```
while mag(ball.pos-paths[i-1]) < d and endPoints[1].pos.y < endPoints[0].pos.y:
                ball.v += norm(paths[i] - paths[i-1]) * sqrt(2 * 9.8 * endPoints[0].pos.y-ball.pos.y)
                ball.pos += ball.v * dt
```

Teachers can extrapolate this to demonstrate that the right curve must satisfy Snell's law $\frac{\sin\theta_1}{v_1} = \frac{\sin\theta_2}{v_2}$ at every tangent. Since $v_1$ and $v_2$ are simply a function of height, we then have $\frac{\sin\theta_1}{\sqrt{h_1}} = \frac{\sin\theta_2}{\sqrt{h_2}}$. Of course, $\frac{\sin\theta}{\sqrt{h}}$ is the differential equation of a cycloid [9]. Hence, the problem is solved by treating the particle as a photon of light and using Snell's Law, a consequence of the PLA.

Students can select any two points to denote the endpoints of the particle's trajectory. They can then draw an arbitrary number of paths between the two endpoints, with a time and action generated for the paths. Each path also involves an animation with real-time graphs of the particle's Kinetic Energy, Potential Energy and Action. Students may also inquire whether any paths that go backward horizontally may minimize the time. Upon experimenting with the simulation, they may reach the conclusion that the shortest-time path must move forward horizontally and can't fold back on itself [8]. Given enough time and trial-and-error, students will realize that the Goldilocks balance between gaining acceleration and ceding horizontal distance occurs in the case of a Brachistochrone.

**3 Conclusion**
In summary, the three numerical simulations outlined above are designed to address previously-known student difficulties with the Principle of Least Action. All of the simulations share a common characteristic in that they ask the user for boundary conditions instead of initial conditions, thus reinforcing the difference between the PLA and Classical Newtonian Mechanics. Each of the programs focuses on fostering active learning in students via mouse interactivity, keyboard input, VPython visualizations, and real-time graphical plots to engage students. In addition, each of the programs give a different perspective of the PLA, through the reflective and refractive properties of light and through the Brachistochrone problem. By learning experientially, changing parameters, and programming the simulations themselves, students will leave with a better understanding of the Principle of Least Action in Optics, as they implement it themselves in Python.

**4 Acknowledgements**
I am grateful to Dr. Kabat for key comments on the manuscript and supporting me in scientific endeavors for many years, including mentoring me in my first research paper. I'm also grateful to Ms. Berry, whose course in Multivariable Calculus and treatment of curvature inspired the writing of this paper.